\documentclass[prb,nobibnotes,11pt]{revtex4}
\usepackage[T1]{fontenc}
\usepackage{revsymb}
\usepackage{graphicx}
\usepackage{amsmath}
\usepackage{psfrag}
\usepackage{color}

\usepackage{setspace}
 
\newcommand{\pphi}{\varphi}

 \newcommand{\cl}{\centerline}
\newcommand{\bi}{\begin{itemize}}
\newcommand{\ei}{\end{itemize}}
\newcommand{\ben}{\begin{enumerate}}
\newcommand{\een}{\end{enumerate}}

\newcommand{\bfi}{\begin{figure}[hbtp]}
\newcommand{\efi}{\end{figure}}
 
\newcommand{\dr}{\partial}
\newcommand{\beq}{\begin{equation}}
\newcommand{\eeq}{\end{equation}}
\newcommand{\beqar}{\begin{eqnarray}}
\newcommand{\eeqar}{\end{eqnarray}}

\begin{document}

\title{Note on dust trapping in inviscid vortex pairs with unequal strengths.}
\author{Tatiana Nizkaya, Jean-Régis Angilella\footnote{Corresponding author~: jean-regis.angilella@ensg.inpl-nancy.fr} \& Michel Buès}

 \affiliation 
{Nancy-Universit\'e, LAEGO, Ecole Nationale Supérieure de Géologie,
 rue du Doyen Roubault, 54501 Vand\oe uvre-les-Nancy, France}

\vskip.5cm 
\cl{Submitted to Physics of Fluids, \today}
\vskip.5cm
\begin{abstract}
{  
We investigate theoretically the motion of tiny heavy passive particles transported in a plane inviscid flow consisting of two   point vortices, in order to understand particle dispersion and trapping during vortex interaction. 
In spite of their large density,
 particles are not necessarily centrifugated away from vortices. It is observed
that they
can have various equilibrium positions in the reference frame rotating with the vortices,   provided the particle response time and the vortex strength ratio lie in appropriate ranges. A stability analysis reveals that some of these points can be asymptotically stable, and can therefore trap particles released in their basin of attraction.
A complete trapping diagram is derived, showing that any vortex pair can potentially become a dust trap, provided the vortex strength ratio is different from 0 (single vortex) and -1 (translating symmetrical vortices).
 Trapping exists for both co-rotating or contra-rotating vortex pairs. In the latter case, particle trapping on a limit cycle is also observed, and confirmed by using Sapsis \& Haller's method [Chaos, 20, 017515, 2010] generalized to non-inertial reference frames. 
}
\end{abstract}
 
\maketitle
   
%% keywords here, in the form: keyword \sep keyword

%% PACS codes here, in the form: \PACS code \sep code

%% MSC codes here, in the form: \MSC code \sep code
%% or \MSC[2008] code \sep code (2000 is the default)
\vskip1cm

\section{Introduction}
  
Dust transport in vortical flows is a topic of interest in many research areas, since many natural or industrial fluid flows carry tiny solid objects. The understanding of particle dispersion, especially in turbulent flows, 
  has been an active field of research in the last century
 (see for example Csanady \cite{Csanady1963},
 Snyder \& Lumley \cite{Snyder1971},  Maxey \cite{Maxey1987}). One of the most 
striking features of these particles, however, is their 
tendency to accumulate in various zones of the flow domain. This behavior has been observed also when particles are advected  in a laminar flow~: particles can be trapped by an attractor, whether chaotic or not, and remain there for long times (Rubin, Jones \& Maxey \cite{Rubin1995} ;  Vilela \& Motter \cite{Vilela2007} ; Haller \& Sapsis \cite{Haller2008}).
This makes the dynamics of inertial particles  significantly different from the one of perfect tracers~: in  divergence-free flows the latter never accumulate.

Accumulation is   rather unexpected for heavy particles in vortical flows. Indeed, these particles are generally centrifugated away from vortical zones
 under the effect of their inertia. In this paper we show that heavy particles can be trapped by vortex pairs, provided the particle response time and the vortex strength ratio lie in an appropriate range.

Inviscid co-rotating point vortex pairs are known to rotate around each other
 under their mutual influence, and
 heavy particles released in such a flow do not have trivial trajectories.
It was shown in a previous work (Angilella \cite{Angilella2010}) that particles released in a vortex pair with identical strengths can have 5 equilibrium positions in the reference frame rotating with the vortex pair, and that two of these positions are asymptotically stable~: they can attract particles. These equilibrium positions result from the balance between the hydrodynamic force, which drives the particle towards the center point of the vortex system,  and the centrifugal force which thrusts the particles outward. In the following sections, particle motion in a vortex pair with unequal strength is examined for any vortex strength ratio. The flow is described in section
\ref{intro}, then particle trapping in equilibrium points is investigated
(section \ref{trappcond}). In section \ref{limitcycl} we show that particles in counter-rotating vortices  can also be trapped by a limit cycle. Particles with
a very large response time are discussed in section \ref{LargeTau}.

 \begin{figure}[htbp]
	\centering
		\includegraphics[width=0.6\textwidth]{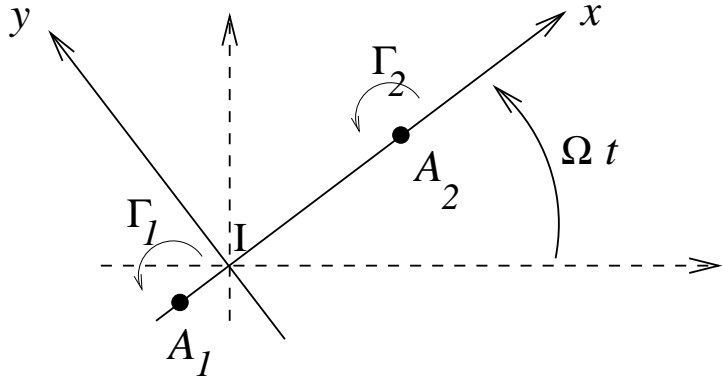}
	\caption{Coordinate system and co-rotating frame (solid lines). The dashed axes are attached to the inertial reference frame.}
	\label{coords}
\end{figure}

\section{Flow and particle motion equations in the co-rotating frame}
\label{intro}

We consider an  incompressible and inviscid flow generated by a system of two point vortices with circulations $\Gamma_{1}$ and $\Gamma_{2}$, located at $A_1$ and $A_2$. 
Elementary vortex dynamics shows that the distance $2d$ between the vortices remains constant, and that they rotate with a constant angular velocity $\Omega=\left(\Gamma_1+\Gamma_2\right)/(8\pi)$ around a point $I$, satisfying  
$\Gamma_1\overrightarrow{IA_1}+\Gamma_2\overrightarrow{IA_2} = \vec 0$,
provided $\Gamma_1 + \Gamma_2 \not = 0$. 
In the reference frame rotating with the vortex pair the flow is steady, and 
vortices remain fixed.
In non-dimensional form,  using $1/\Omega$ for time  and $d$ for lengths, the vortex positions in this frame are~:
$$
\vec{x}_1=\left(-\frac{2}{\gamma+1},0\right), \hspace{.2cm}\vec{x}_2=\left(\frac{2\gamma}{\gamma+1},0\right),\hspace{.2cm}\gamma=\Gamma_1/\Gamma_2
$$
where the coordinate system $(x,y)$ is centered at $I$, with $Ix$ parallel to $A_1 A_2$
(Fig. \ref{coords}).
The fluid velocity field reads 
 $\vec{V}_f(\vec x,\gamma)=\left(\frac{\partial \Psi}{\partial y}, -\frac{\partial \Psi}{\partial x}\right),
$
where $\Psi$ is the streamfunction  in the co-rotating frame which can be written as~: 
 $$
\Psi(\vec{x})=\gamma\Phi(\vec{x}-\vec{x}_1)+\Phi(\vec{x}-\vec{x}_2)+|\vec{x}|^2/2,
$$
with $\Phi(\vec{x})=-\frac{2}{1+\gamma}\ln\left|\vec{x}\right|.$
Vorticity is constant and equals $-2$ everywhere, except at the vortex centers.
Figs.   \ref{cloudsco} and  \ref{cloudscontra}  shows typical relative streamlines 
$\Psi = constant$
for various strength ratios $\gamma$. We will limit ourselves to $\gamma\in ]-1,1]$, since the
case  $|\gamma| > 1$ can be treated by exchanging the role of vortices, i.e. by setting $\gamma \to 1/\gamma$.

 Particles are assumed to be non-interacting, non-brownian, much heavier than the fluid, and with a low Reynolds number (based on their radius and on the slip velocity).  Clearly, the first approximation is no longer valid when particles tend to accumulate (see for example Medrano et al. \cite{Medrano2008}). However, we will use it for the sake of simplicity in this paper, and keep the effect of particle interactions as a perspective to this work. 
The simplest  motion equation of these  heavy particles at low Reynolds number reads,
in non-dimensional form~:
% \beq
% m_p \frac{d^2 \vec{X}_p}{dt^2} =  6 \pi \mu a \Big(\vec{V}_f - \frac{d \vec{X}_p}{dt} \Big) - m_p \vec \gamma_e - m_p \vec \gamma_c
% \label{eqmvt0}
% \eeq
\beq
\tau \frac{d^2 \vec{X}_p}{dt^2} =    \vec{V}_f(\vec X_p) - \frac{d \vec{X}_p}{dt}   +
\tau \Big(\vec X_p - 2 \vec e_z \times \frac{d \vec{X}_p}{dt} \Big)
\label{eqmvt1}
\eeq
where $\vec X_p(t) = (x_p,y_p)$ denotes the particle position at time $t$,   $\vec e_z$ is the unit vector along the $z$ axis, 
and 
$\tau$ is the non-dimensional response-time of the particle (Stokes number). For spherical particles with mass $m_p$ and radius $a$ the Stokes
number reads $\tau =  \Omega m_p/(6 \pi \mu a)$, where $\mu$ is the fluid viscosity.  
The last two terms in equation (\ref{eqmvt1}) are the   (non-dimensional)
centrifugal  and  Coriolis pseudo-forces respectively. 
Because particles are much heavier than the fluid, Eq. (\ref{eqmvt1}) is valid even 
if the Stokes number $\tau$ is not small, as added mass force, pressure gradient of the undisturbed flow, lift and Basset force are negligible.
% \section

\section{Trapping condition.} 
\label{trappcond}

Introducing a new vector variable
$\vec{z}=(x_p,y_p,\dot{x}_p,\dot{y}_p),$
  equation (\ref{eqmvt1}) can be rewritten as a system of first order equations: 
\begin{equation}\label{system}
\left\{\begin{array}{l}
\dot{z}_1=z_3\\
\dot{z}_2=z_4\\
\dot{z}_3=-z_3/\tau+z_1+2z_4+u_x(z_1,z_2,\gamma)/\tau\\
\dot{z}_4=-z_4/\tau+z_2-2z_3+u_y(z_1,z_2,\gamma)/\tau,\\
\end{array}
\right.
\end{equation}
where $\vec V_f = (u_x,u_y)$. In vector form this set of equations writes $d{\vec{z}}/dt=\vec{F}(\vec{z},\gamma,\tau)$.
To find the conditions of particle trapping, we search for attracting equilibrium points of (\ref{system}) :
$
\vec{F}(\vec{z},\gamma,\tau)=\vec 0.
$ 
If   all the eigenvalues  of $\nabla \vec{F}$ have strictly negative real parts, then  $\vec X_{eq}$
is an asymptotically stable hyperbolic point.
It can be shown (see Appendix \ref{appA}) that the eigenvalues of $\nabla\vec{F}$ at the
 equilibrium point only depend 
on the jacobian $J = u_{x,x} u_{y,y} - u_{y,x} u_{x,y}$ 
 (the comma indicates spatial derivation)  of the velocity field at this point, and the stability criterion is $-\tau^2<J<1$.
For our velocity field the condition $J<1$ is  automatically satisfied. 
% This means that particles with response time $\tau$ can be trapped in a vortex system with  strength ratio $\gamma$ if and only if there exists a point 
Let $\vec{X_{eq}}$  be an equilibrium point : 
\begin{equation}
\label{trapping}
\vec{f}(\vec{X}_{eq},\gamma,\tau)=\vec{X}_{eq}+\vec{V}_f(\vec{X}_{eq},\gamma)/\tau=0.
\end{equation}
By virtue of the implicit functions theorem,  $\vec{X}_{eq}$ depends smoothly on $(\gamma,\tau)$ unless $\mbox{det} (\nabla\vec{f}\left|_{X_{eq}}\right. )=0$. Because $\mbox{det} (\nabla\vec{f}\left|_{X_{eq}}\right. )=J/\tau^2+1$, this
condition coincides exactly with the boundary of the stability domain.
% Suppose that such a point $\vec{X}_{eq}$ exists for some $(\gamma_0,\tau_0)$. 
% It can be shown that $\vec{X}_{eq}$ depends smoothly on $\gamma,\tau$, unless $\nabla\vec{f}\left|_{X_{eq}}\right.=J/\tau^2+1=0$ (which coincides exactly with the stability boundary above). 
Hence, the point $\vec{X}_{eq}$ will exist and remain stable until $(\gamma,\tau)$ reaches some critical value defined by the system of equations:
\begin{equation}
\label{trapping2}
\left\{\begin{array}{l}\vec{X}_{eq}+\vec{V}_f(\vec{X}_{eq},\gamma)/\tau=0\\J(\vec{X}_{eq},\gamma)+\tau^2=0.\end{array}\right.
\end{equation}
This is a system of 3 equations for 4 scalar variables $(\vec{X}_{eq},\tau,\gamma)$, which defines a family of curves in a four-dimensional  space.
The projection of these curves on the $(\tau, \gamma)$-plane splits it  into subdomains, within which the number (and stability properties) of equilibrium points does not change. It is therefore sufficient to investigate the stability of equilibrium points for just one set of parameters $(\gamma,\tau)$ from each subdomain to get the full picture.

\begin{figure}[htbp]
	\centering
		\includegraphics[width=0.7\textwidth]{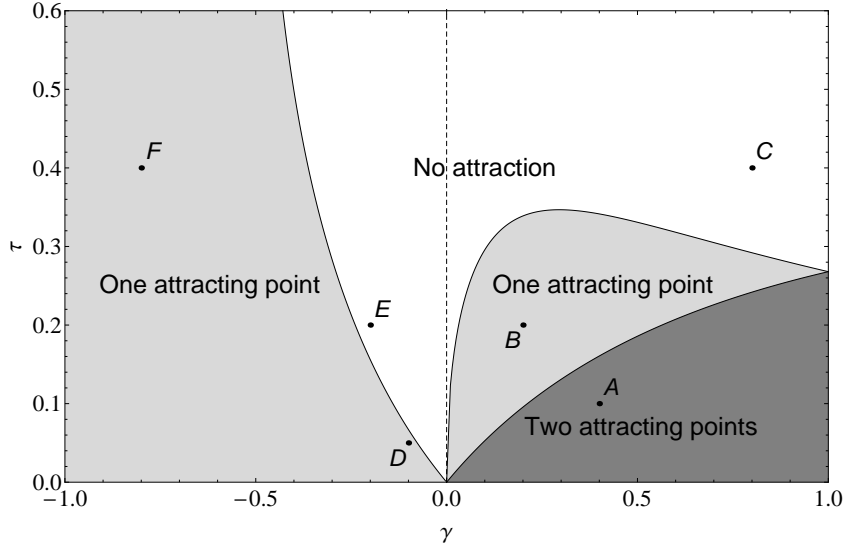}
	\caption{Particle trapping diagram.}
	\label{diagram}
\end{figure}

To find these critical curves we took a grid covering $\gamma\in ]-1,1]$ and for each $\gamma$ solved the system (\ref{trapping2}) numerically. The resulting trapping diagram is shown in Fig.\ref{diagram}, for $ \gamma \in ]-1,1]$ and $0 < \tau < 0.6$. (Trapping also exists for larger $\tau$ : particles with a large response time are discussed in the last section.) 

\begin{figure}
\includegraphics[width=0.4\textwidth]{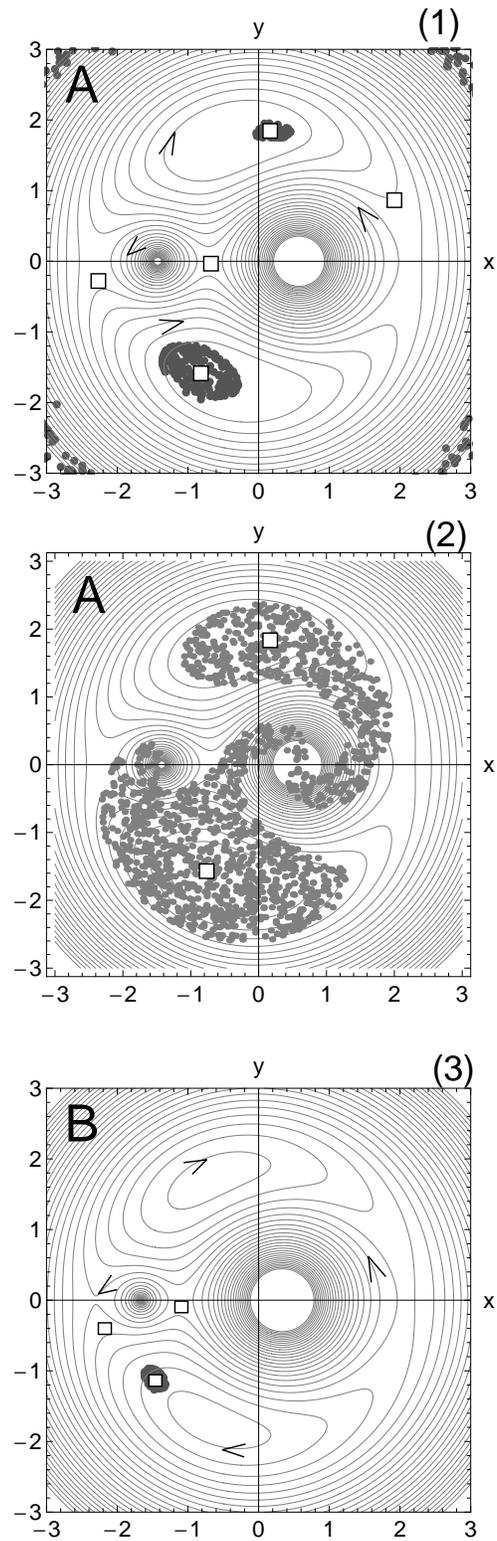}  
\caption{Flow streamlines, particle clouds (black points) and equilibrium positions (white squares) for co-rotating vortices with different sets of parameters~:
A: $\gamma=0.4$, $\tau=0.1$ ; B: $\gamma=0.2$, $\tau=0.2$.
Grey dots in Fig.  (2)   indicate the initial position of trapped particles in case A.
}
\label{cloudsco}
\end{figure}

\begin{figure}
\includegraphics[width=0.4\textwidth]{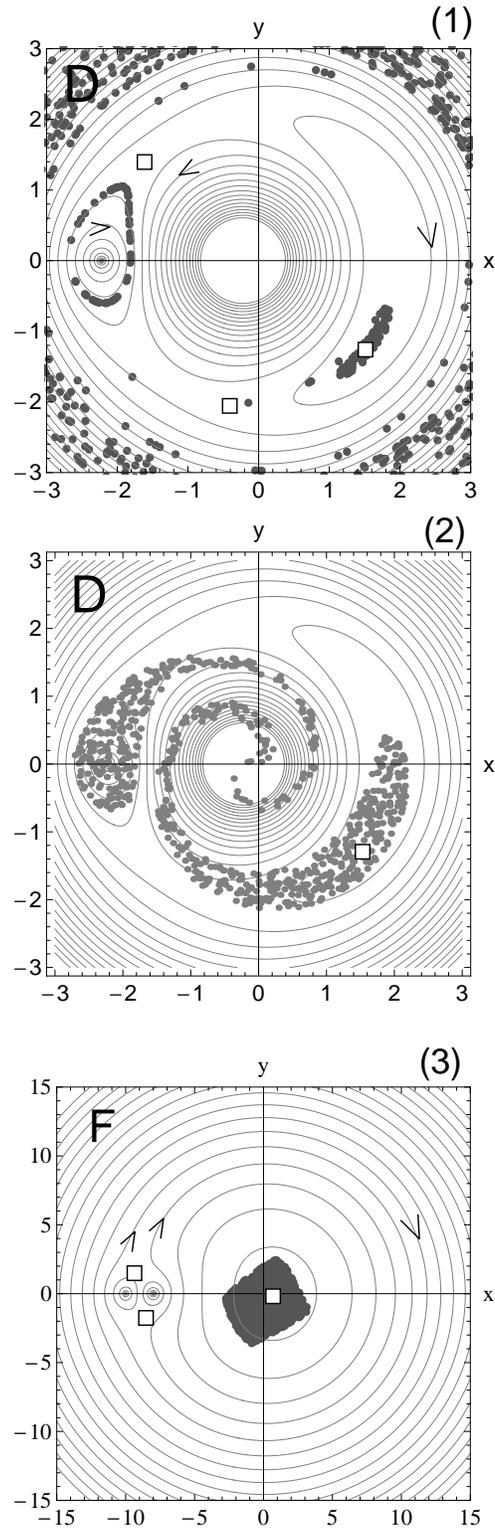}
\caption{Flow streamlines, particle clouds (black points) and equilibrium positions (white squares) for contra-rotating vortices with different sets of parameters~:
D: $\gamma=-0.1$, $\tau=0.05$ ;  F: $\gamma=-0.8$, $\tau=0.4$.
Figure F has been rescaled to account for the large radii of vortices trajectories when $\gamma$
is close to -1. Grey dots in Fig.  (2)   indicate the initial position of trapped particles in case D.
}
\label{cloudscontra}
\end{figure}

As expected, for $\gamma=1$ the critical value of $\tau$ coincides with the one obtained analytically for identical vortices \cite{Angilella2010} : $\tau=2-\sqrt{3}$.
Numerical simulations of particle clouds have been run for the sets of parameters (A,B,C,D,E,F) marked on figure \ref{diagram}.
At $t=0$ particles were randomly distributed in a square $[-3,3]\times[-3,3]$.
Two types of initial velocities were considered : (i) no-slip initial condition ($\dot {\vec X}_p(0) = \vec V_f(\vec X_p(0))$), and (ii) particles at rest in the non-rotating frame ($\dot {\vec X}_p(0) =  - \vec \Omega \times \vec X_p(0)$). In these runs, we observed that for both (i) and (ii) the qualitative behavior of  particles was the same, and the percentage of trapped particles did not change significantly. (Note however that particles with a large initial velocity might be located outside the 4-dimensional basin of attraction of the equilibrium point(s), and 
could therefore avoid trapping.  )
% $$
% \begin{array}{lll}
% A: \gamma=0.4, \tau=0.1 & B: \gamma=0.2, \tau=0.2 & C: \gamma=0.8, \tau=0.4\\
% D: \gamma=-0.2, \tau=0.05 & E: \gamma=-0.2, \tau=0.2 & F: \gamma=-0.8, \tau=0.4
% \end{array}
% $$
Figs.   \ref{cloudsco} and  \ref{cloudscontra}  present the resulting particle clouds (black dots) at $t=30$ in cases  A, B, D,  and at $t=500$ for F (particles take much longer time to converge in this case),  by using the no-slip initial condition. 
We observe that some particles converge to the
 attracting points predicted by the diagram, as expected.
Grey dots, in Figs.  \ref{cloudsco}(2) and  \ref{cloudscontra}(2),  indicate the initial position of trapped particles (this is a cut of the basin of attraction). Clearly, the typical size of this basin is of order unity so that the percentage of trapped particles is non-negligible, about 15-20 \% in cases A and D. 
We have checked that no trapping occurs in cases C and E, in agreement with the trapping diagram of Fig. \ref{diagram}.

These results show that any vortex pair can trap particles, provided the vortex strength ratio $\gamma$  differs from -1 or 0.

\section{Particle trapping into a limit cycle.} 
\label{limitcycl}

Figure \ref{cloudscontra}(1) suggests that, apart from the attracting points, a limit cycle could exist for small $\tau$ and relatively small negative $\gamma$.
The existence of a limit cycle  can be predicted by using a criterion by   Sapsis and   Haller  \cite{Haller2010}.
To adapt this criterion to our equations we repeat the procedure from these authors, using an asymptotic expression of the particle velocity in powers of $\tau$. 
This procedure leads us to the following inertial equation :
$$
\frac{d\vec{X}_p}{dt}=\vec{V}_f(\vec{X}_p) + \tau \vec V_1(\vec{X}_p) + O(\tau^2)
$$
with
$ \vec V_1(\vec{X}_p) = 
  \vec{X}_p -2\vec{e}_z\times  \vec{V}_f  - \nabla \vec{V}_f . \vec{V}_f  .
$
With the use of this equation  we get the following necessary condition for the existence of a limit cycle $\tau$-close to a flow streamline $S_0$ (defined by $\Psi(x,y)=\Psi_0$)~:
\beq
I(S_0)=\int\limits_{S_0}  \vec V_1 \cdot \vec{n}dS=0 \quad\mbox{and}\quad
 \int\limits_{S_0}  \vec\nabla(\vec V_1\cdot \vec n)  \cdot \vec{n}dS <0 ,
\label{crit}
\eeq
where $\vec{n}$ is the outer normal to $S_0$.
The former equation in (\ref{crit}) is a consequence of the existence of an invariant manifold for the particle dynamics $\tau$-close to $S_0$. The latter
manifests the fact that this manifold is attractive.

\begin{figure}
\includegraphics[width=0.5\textwidth]{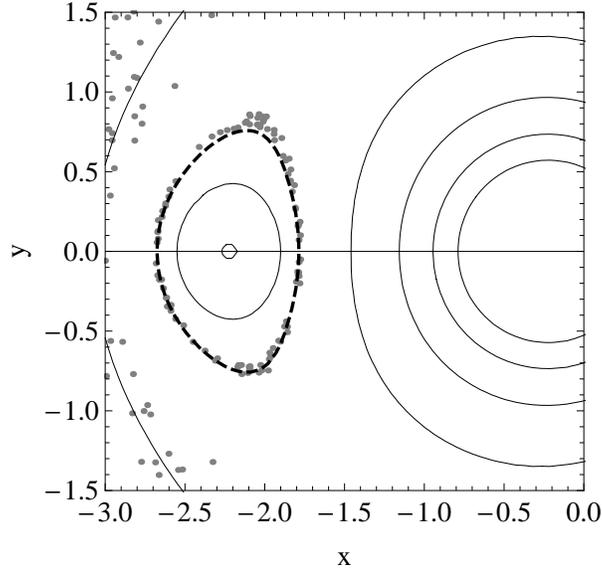}
\caption{Particle cloud at $t=50$ for $\gamma=-0.1, \tau=0.01$ (grey points) compared with the predicted limit cycle (dashed black line).}
\label{clouds2}
\end{figure}

Now the problem is to find a value $\Psi_0$ such that 
$I(S_0(\Psi_0))=0.$
This is  a one-dimensional root-finding problem, which can be solved by applying
 the simplest bisection method, which does not require any additional information on $I$.
The value of $I$ at any given $\Psi$ is obtained by numerical integration along the corresponding streamline.
For $\gamma=-0.1$ this procedure converges to the value of $\Psi_0\approx-0.761$.
The simulation carried out for particles with $\tau=0.01$ confirms
 that the limit cycle indeed
lies quite close to this streamline (Fig. \ref{clouds2}).

 \section{Particles with a large response time}
\label{LargeTau}

\begin{figure}
\includegraphics[width=0.8\textwidth]{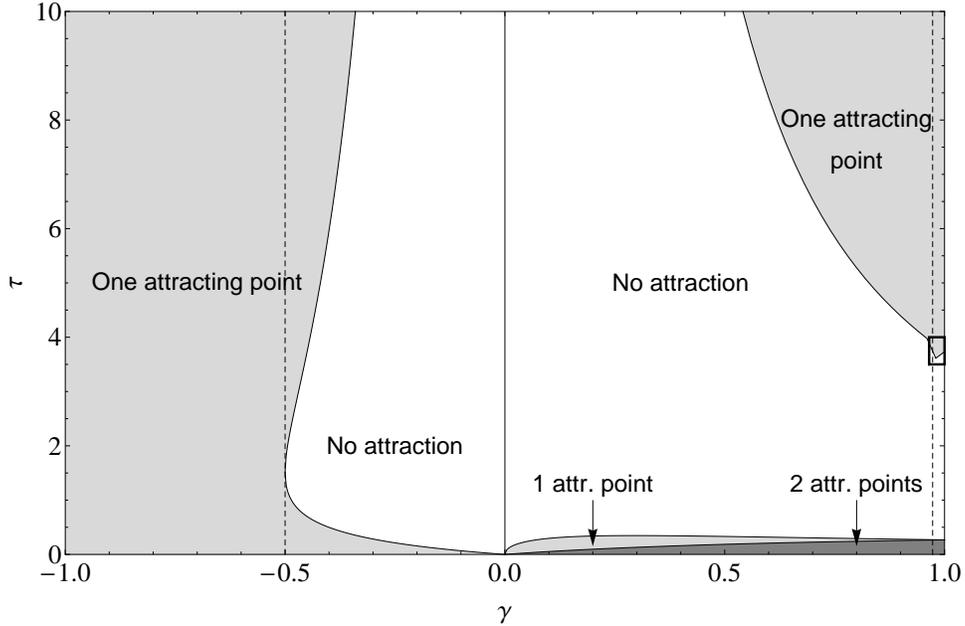}
\caption{Full trapping diagram up to $\tau=10$. }
\label{full}
\end{figure}

 Figure \ref{full} shows the complete trapping diagram up to $\tau = 10$ : it suggests that particle trapping exists in the limit of large $\tau$.
Indeed,   the upper critical curves in this diagram have a vertical asymptote at $\gamma\rightarrow 0$, that is  $\tau\approx 2/\gamma^2$ (see Appendix \ref{appB}) : the grey zones are not bounded from above.
Therefore, a stable equilibrium point can exist no matter how heavy the particles are. Trapping is unexpected for particles with a 
large $\tau$ because
 their trajectories are ballistic : they are weekly influenced by the fluid. However, there is no contradiction here. The equilibrium point in coordinate-velocity space $\vec z_{eq}=(x_{eq},y_{eq},0,0)$ is stable only locally and the basin of attraction for this point can be very small. Indeed,  particles injected with a zero initial velocity (in the rotating frame) in the vicinity of the equilibrium point,  will be slowly pushed to the attractor by the flow (with a rate of order $1/\tau$, as shown in Appendix \ref{appB}). In contrast, particles injected with a finite velocity will continue their inertial motion and will therefore avoid trapping.

There are  some  points on the diagram where the derivative of the critical curves is no longer defined. These are the bifurcation values of $\gamma$, i.e. the values that turn the determinant of the system (\ref{trapping2}) at the solution point to zero :
$\det{\left(\frac{\dr \vec{g}}{\dr \vec{p}}\right)}=0,$ 
where $\vec{g}(\vec{p})$ is the right-hand side of the system (\ref{trapping2}), $\vec{p}=(x_{eq},y_{eq},\tau)$ is an extended vector of parameters.
Solving this equation for $\gamma$ together with the system (\ref{trapping2}), we get the following values :
$\gamma_1\simeq -0.50, \hspace{1ex} \gamma_2 \simeq 0.97$. We can see from the diagram that 
contra-rotating vortices with $\gamma<\gamma_1$ can (in principle) trap particles with any $\tau$, while for each $\gamma>\gamma_1$ there exists a range of particles that are never trapped.

The second point $\gamma_2$ corresponds to the appearance of two more branches of the critical curve at $\tau \simeq 3.56 $, but numerical simulations show that the percentage of trapped particles in these conditions is neglible (around 1\%), so this part of the diagram is out of our interest.

\newpage
\section{Conclusion} 

We have shown that inviscid vortex pairs can permanently trap heavy particles, provided the vortex strength ratio and the particle response time lie in the grey zones of the trapping diagram shown in Fig.  \ref{diagram}. Trapping can occur in the vicinity of various attracting points rotating with the vortex pair, where the centrifugal force balances the Stokes drag, for both co-rotating or contra-rotating vortices. The trapping diagram shows that any vortex pair can potentially become a dust trap, provided the vortex strength ratio is different from 0 (single vortex) and -1 (translating symmetrical vortices). 
Contra-rotating vortices with $\gamma<\gamma_1 \simeq 0.5$  can   trap particles with any $\tau$. In contrast, for $\gamma>\gamma_1$ there exists a range of particles that are never trapped. Numerical simulations show that the percentage of trapped particles with small or moderate $\tau$ can reach tens of percents. For  particles with a large inertia ($\tau \gg 1$) the convergence rate is very slow ($\sim 1/\tau $) and the percentage of trapped particles is negligible.  

In the   case of counter-rotating vortices with small $\tau$ and relatively small $\gamma$, a limit cycle was found  by using Sapsis \& Haller's method \cite{Haller2010}. A similar particle behavior has been observed in a cellular vortex flow model used for the analysis of the solar nebula (Tanga {\it et al.}  \cite{Tanga1996}). 

An important issue is the effect of viscosity (which smoothes the fluid velocity field near the vortex cores and finally leads to vortex merging) on the trapping phenomenon. It could be natural to expect some sort of trapping provided  the particle trapping time is much less than the characteristic time of vortex diffusion. Particle trapping in the viscous case has indeed been observed numerically  \cite{Angilella2010} for identical vortices. However, particle trapping by a viscous vortex pair with any strength ratio is still  an open question.

% In the viscous case trapping was observed too\cite{Angilella2010}. The effect of viscosity  (which smoothes the fluid velocity near the vortex cores and finally leads to vortex merging) on particle trapping  is still an open question. 
% \begin{figure}
% \includegraphics[width=0.7\textwidth]{fork1}
% \caption{ }
% \label{fork1}
% \end{figure}

 Trapping leads to large collision rates and  could be of interest to understand the detailed dynamics of 
various particle-laden flows where vortex pairing is known to occur. For example, anti-cyclonic vortices in protostellar disks are known to trap particles, with the help of the Coriolis force due to the disk rotation (Barge \& Sommeria \cite{Barge1995}, Tanga {\it et al.} \cite{Tanga1996}, Chavanis \cite{Chavanis2000}). 
The analysis of the behavior of particles during the pairing of anticyclonic vortices could therefore be a topic of interest.  

 \appendix
% 
% \section
\section{Stability condition of the equilibrium points.}
\label{appA}

Consider the characteristic equation of (\ref{system}), taking into account the properties of the flow (vorticity equals -2, divergence equals 0):
$$\lambda^4+\frac{2}{\tau}\lambda^3+\left(2+1/\tau^2\right)\lambda^2+1+J/\tau^2=0,$$
where $\lambda$ denotes any eigenvalue of $\nabla\vec F$.
Using the change of variables $\zeta=\lambda+1/2\tau$, we obtain a biquadrate equation:
$$16\tau^4\zeta^4+8\tau^2(4\tau^2-1)\zeta+8\tau^2(2\tau^2+2J-1)+1 = 0$$
which has the following roots:
$\zeta_{\pm}^2=({1-4\tau^2})/{4\tau^2}\pm\sqrt{-J}/\tau$.
Consider two possible cases: $J<0$ and $J>0$.

When $J<0$, $\zeta^2_{\pm}$ are both real and  $\zeta^2_{+}>\zeta^2_{-}$.
Now if $\zeta^2_{\pm}<0$, all the roots $\zeta_{\pm}$ are imaginary, so $\Re[\lambda]=-1/{2\tau}$ and the equilibrium point is stable. If at least $\zeta^2_{+}>0$, then  $\zeta^{(1,2)}_{+}=\pm\sqrt{\zeta^{2}_{+}}$ and $\max \Re[ \lambda]=\sqrt{\zeta^2_{+}}-1/2\tau$. After some simple manipulations we find that $\max\Re[ \lambda]>0\Leftrightarrow J<-\tau^2.$

When $J>0$ the roots of the biquadrate equation are complex-conjugate:
$\zeta^2_{\pm}=1/4\tau^2-1\pm i \sqrt{J}/\tau=a\pm i b$.
Using trigonometric representation and the Moivre formula, we find that
$
\zeta^{(1)}_{\pm}=r^{1/2}(\cos{\pphi/2}\pm i\sin{\pphi/2})
$
and $
 \zeta^{(2)}_{\pm}= - \zeta^{(1)}_{\pm}$,
% $$
% \zeta^{(2)}_{\pm}=-r^{1/2}(\cos{\pphi/2}\pm i\sin{\pphi/2}),
% $$
where $r=\sqrt{a^2+b^2},\cos{\pphi}=a/r$,
so $\max{\Re[\lambda]}=|r^{1/2}\cos{\pphi/2}|-1/{2\tau}.$
Using the fact that $\cos^2{\pphi/2}=(1+\cos{\pphi})/2$ we find that $\max\Re[ \lambda]>0\Leftrightarrow J>1.$
 Combining these results, we obtain the stability criterion :
the equilibrium point $\vec{X}_{eq}$ is stable if and only if
$-\tau^2<J(\vec{X}_{eq},\gamma)<1$.

\section{Asymptotic behaviour for   large $\tau$}
\label{appB}
 
First we consider the case of finite $\gamma\ne0,-1$, taking $\varepsilon=1/\tau\ll 1$ as a small parameter, and looking for the equilibrium point in the following form:
\beq \label{eq1}x_{eq}=C_x\,\varepsilon^m,\hspace{1ex} y_{eq}=C_y \, \varepsilon^k.\eeq
Injecting this ansatz into the velocity field and keeping only the leading order terms in the limit $\varepsilon \to 0$, we obtain  :
$$\begin{array}{ll}
u_x\simeq C_y\varepsilon^k\left(1-\frac{(1+\gamma)(1+\gamma^3)}{\gamma^2}\right)\\
u_y\simeq 2\left(\gamma-\frac{1}{\gamma}\right)
\end{array}
$$
The equilibrium equations (\ref{trapping}) then read (to leading order):
$$
\begin{array}{ll}
C_y\varepsilon^{k+1}\left(1+\frac{(\gamma+1)(\gamma^3+1)}{\gamma^2}\right)+C_x\varepsilon^m=0,\\
2(\gamma-1/\gamma)\varepsilon+C_y\varepsilon^k=0.
\end{array}
$$ 
To ensure the consistency of these equations in the limit $\varepsilon\rightarrow0$ we have to set $m=2,k=1$. This leads to a system of two equations for $C_x$ and $C_y$, which can be readily solved :
$$
\begin{array}{ll}
C_y=-2(1-1/\gamma),\\
C_x=2(1-1/\gamma)(1-B),
\end{array}
$$ 
with $B={(\gamma+1)(\gamma^3+1)}/{\gamma^2}$.
Therefore, an equilibrium point of the form (\ref{eq1}) does exist for large $\tau$.
To investigate its stability we calculate the Jacobian of the fluid velocity field at this point : $J=(1-B^2)+O(\varepsilon)$.
Using the expression for the roots of the characteristic equations from App. \ref{appA} and the Moivre formula, we obtain the following leading order estimates for the real part of the eigenvalues :
$$\Re[\lambda]=\left\{
\begin{array}{ll}
\varepsilon/2(-1\pm\sqrt{1-B^2}),&B^2<1,\\
-\varepsilon/2,&B^2\geq1.
\end{array}\right.$$
Evidently, $\max\Re[\lambda]<0$, so this equilibrium point is always (locally) stable for large $\tau$. The convergence rate is, however, very slow for these heavy particles and decays as $\tau^{-1}$.

The asymptotics above doesn't work for $\gamma\rightarrow 0$. To investigate this case we consider $\gamma$ as a small parameter and search for the equilibrium point in the following form : $  x=C_x\gamma^m,\hspace{1ex} y=C_y\gamma^m.$ 
The same method as above leads to $\varepsilon=A\gamma^n$ with $n=2$ and $m=1$,
and the corresponding solution is
$C_x=1\pm\sqrt{1-4A^2}$ and $C_y=2A$. Clearly, $C_x,C_y$ are real only if $A\leq1/2$. The value $A=1/2$ corresponds to the upper critical curves in the $(\tau,\gamma)$-plane, that is $\tau=2/\gamma^2$ (we have checked that the two upper curves of the numerical trapping diagram do coincide with this asymptote when $\gamma\rightarrow0$). When $A<1/2$ two equilibrium points exist, corresponding to $C_x=1+\sqrt{1-4A^2}$ and $C_x=1-\sqrt{1-4A^2}$. For the former point the roots of the biquadrate equation (Appendix \ref{appA}) read :
$$\zeta_{\pm}=-1\pm\frac{1}{1+\sqrt{1-4A^2}}.$$
These roots are always real and negative, so $\Re[\lambda]=-\frac{1}{2\tau}$. This means that  the point is locally stable, with a slow convergence rate  (as in the case above). Similarly, one can check that the latter point is always unstable in the limit $\tau\rightarrow\infty$, with a divergence rate of order unity.

%\bibliography{../../../Biblio/global}

% \listoffigures

\end{document}